\documentclass[intlimits,twoside,a4paper]{article}

\usepackage{graphicx}
\usepackage[T2A]{fontenc}
\usepackage[cp1251]{inputenc}

\usepackage{cmpj2}

\issue{2013}{16}{4}{43001}
\doinumber{10.5488/CMP.16.43001}

\title{Nonmonotonous pressure as a function of the density in a fluid without attractive forces}

\author[D.~Henderson]{D.~Henderson\thanks{E-mail: doug@chem.byu.edu}}

\authorcopyright{D.~Henderson, 2013}

\address{Department of Chemistry and Biochemistry,
Brigham Young University, Provo UT 84602--5700}

\date{Received April 10, 2013, in final form May 18, 2013}

\begin{document}
\maketitle


\begin{abstract}

A simple result for the pressure of a hard sphere fluid that was developed many years ago by Rennert is extended in a straightforward
manner by adding the terms that are of the same form as the Rennert's formula.  The resulting expression is  moderately accurate but
its accuracy does not necessarily improve as additional terms are included.  This expression has the interesting consequence that the
pressure can have a maximum,
as the density increases, which is consistent with the freezing of hard spheres.  This occurs solely as a consequence of repulsive
interactions.  Only the Born-Green-Yvon and Kirkwood theories show such a behavior for hard spheres and they require a numerical solution of
an integral
equation.  The procedure outlined here is {\it ad hoc} but is, perhaps, useful just as the popular Carnahan-Starling equation for the hard
sphere pressure is also {\it ad hoc} but useful.

\pacs 05.20.-y, 05.20.Jj, 64.10.+h, 64.30.+t, 64.70.Hz

\end{abstract}

\section{Introduction}

For a number of years, the author has been intrigued by a result obtained by Rennert~\cite{1.} for a hard sphere fluid.  For this fluid, the
interaction potential, $u(R_{12})$, for a pair of spheres whose centers are located at ${\bf r}_1$ and ${\bf r}_2$, vanishes if the
separation of the
centers, $R_{12}=|{\bf r}_1-{\bf r}_2|$, exceeds their diameter, $d$.  Since the spheres are hard, they cannot overlap and $u(R)$ is
infinite if $R<d$.  To keep the discussion simple, here the hard spheres are all assumed to be of the same diameter.  Hard spheres are
important
because the `structure' of a simple liquid or dense gas, such as argon, is, apart from the small perturbing effect of the attractive
dispersion forces, determined by hard sphere interactions \cite{2.}.

The connection between the pressure of a fluid of $N$ molecules and their interactions is provided by the configurational
partition function, $Q_N$,
\begin{equation}
 Q_N=\frac{1}{N!}\int\exp\left[-\beta U({\bf r}_1\,,\,\dots , {\bf r}_N)\right]\rd{\bf r}_1\,,\,\dots ,\rd{\bf r}_N\,,
\end{equation}
where $\beta=1/kT$, and $k$ and $T$ are the Boltzmann constant and temperature, respectively.
The interaction energy $U$ is assumed to be the pairwise additive sum of the pair potentials,

\begin{equation}
 U({\bf r}_1,\dots ,{\bf r}_N)=\sum_{i<j=1}^Nu(r_{ij}).
\end{equation}

The challenge is to determine the configurational partition
function, since once it is known, the density dependent part of the Helmholtz function, $A$, and the pressure, $p$, for a given value of
the density, $\rho=N/V$, where $V$ is the volume, can be obtained from
\begin{equation}
A=-kT\ln Q_N
\end{equation}
and
\begin{equation}
p=-\frac{\partial A}{\partial V}\,.
\end{equation}
Since $N$ is enormous (of the order of Avogradro's number), this can be done only approximately.

\section{Rennert's method}

Rennert considers a collection of systems that have $n$ hard spheres, where $n$ is a variable.   He defines
\begin{equation}
q_n=\frac{Q_{N+1}}{VQ_{N}}\,,
\end{equation}
$Q(0)=1$. From this it follows that
\begin{equation}
\frac{p}{\rho kT}=1-\rho\frac{\partial}{\partial\rho}\left[\frac {1}{N}\int_0^N\ln q_n\rd n\right].
\end{equation}

Reasoning by analogy to one dimension, Rennert suggests the ansatz
\begin{equation}
 \ln q_n=\left(1-\frac{nv_d}{V}\right)-\epsilon\frac{nv_d}{V-nv_d}
\end{equation}
from which, after some algebra, it follows that
\begin{equation}
\frac{p}{\rho kT}=\frac{\epsilon}{1-\rho v_d}-\frac{1-\epsilon}{\rho v_d}\ln(1-\rho v_d),
\end{equation}
where $v_d$ and $\epsilon$ are the parameters to be chosen.  In one dimension, Rennert chooses $v_d=d=b$,
where $B_2=b$ is the correct second virial coefficient in one dimension, and $\epsilon=0$.  The choice $v_d=d$ is sensible because for
$\rho d=1$, the hard particles fill space and a singularity is to be expected.  The virial coefficients for one dimensional hard particles
that result from equation~(8) are
\begin{equation}
 B_n=\frac{1+(n-1)\epsilon}{n}d^{n-1}\,.
\end{equation}
They are correct with $\epsilon=0$.  The resulting equation of state is

\begin{equation}
 \frac{p}{\rho kT}=\frac{1}{1-y},
\end{equation}
which, for hard particles in one dimension, is exact.

Rennert also applied equation~(8) in three dimensions (hard spheres).  With the choices, $v_d=B_2=b=2\pi d^3/3$ and $\epsilon=8\pi/3\sqrt 2-1$,
equation~(8) yields the correct second and third virial coefficients for hard spheres.  As is seen in figure~1, Rennert's procedure gives
fair results for hard spheres.  In summary, for hard particles, using equation~(8) and choosing $v_b$ so that $\rho_b$ is the density at close
packing and $\epsilon$ yields the correct
second virial coefficient, exact results are obtained in one dimension and fair results are obtained in three dimensions.

\section{Extension}

A reasonable extension of equation~(7) is
\begin{equation}
 \ln q_n=\ln\left(1-\frac{nv_d}{V}\right)-\sum_{i=1}^{\infty}\epsilon_i\left(\frac{nv_d}{V-nv_d}\right)^i,
\end{equation}
which, neglecting the terms in the sum for $i>4$  leads to the following extension of equation~(8)
\begin{equation}
 \frac{p}{\rho kT}=-\frac{1}{\rho v_d}\ln(1-\rho v_d)+\sum_{i=1}^4 a_i\epsilon_i\,,
\end{equation}
where
\begin{equation}
a_1=\frac{1}{\rho v_d}\ln(1-\rho v_d)+\frac{1}{1-\rho v_d}\,,
\end{equation}
\begin{equation}
a_2=-\frac{2}{\rho v_d}\ln(1-\rho v_d)-3\frac{1}{1-\rho v_d}+\frac{1}{(1-\rho v_d)^2}\,,
\end{equation}
\begin{equation}
a_3=\frac{3}{\rho v_d}\ln(1-\rho v_d)+\frac{11}{2(1-\rho v_d)}-\frac{7}{2(1-\rho v_d)^2}+\frac{1}{(1-\rho v_d)^3}\,,
\end{equation}
and
\begin{equation}
a_4=-\frac{4}{\rho v_d}\ln(1-\rho v_d)-\frac{25}{3(1-\rho v_d)}+\frac{23}{3(1-\rho v_d)^2}-\frac{13}{3(1-\rho v_d)^3}+\frac{1}{(1-\rho v_d)^4}\,.
\end{equation}

Keeping only $\epsilon_1$ yields Rennert's result.  Each of the $\epsilon_i$ can be chosen to give $B_{i+1}$, which are known to high
order \cite{2.}.  The result for $B_i$ that follows from the above result has a pleasing form.  It is
\begin{equation}
 B_i/v_b^{i-1}=\frac{1}{i}+\epsilon_1\frac{i-1}{i}+\epsilon_2\frac{(i-1)(i-2)}{i}+\epsilon_3\frac{(i-1)(i-2)(i-3)}{2i}
+\epsilon_4\frac{(i-1)(i-2)(i-3)(i-4)}{6i}+\cdots \,.
\end{equation}
The sum is terminated when a negative term is encountered.
If one wishes to add an additional term in equation~(12), the previously determined values of $\epsilon_i$ remain unchanged.  In one dimension, $
\epsilon_1=1$ and the additional
$\epsilon_i=0$.  In three dimensions, $\epsilon_1=4.92384391$, $\epsilon_2=2.80081908$, $\epsilon_3=-0.918729979$ and $\epsilon_4=-0.218154034$.

\begin{figure}[htb]
\begin{center}
\includegraphics[width=7.0cm,clip]{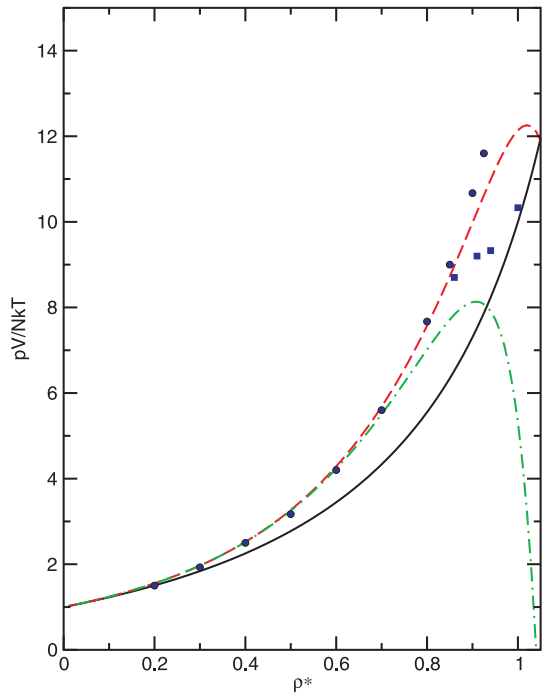}
\caption{(Color online) Pressure of hard spheres as a function of density.  The points give the simulation results taken from reference 2; the circles and squares give the simulation results for the fluid and solid branches, respectively.
The solid curve gives the results obtained from equation~(8).  The dashed and dot-dashed curves give
the results obtained from equation~(12) with $\epsilon_i$ included for $i\leqslant 3$
and $i\leqslant 4$, respectively.}
\end{center}
\end{figure}

Results for $p/\rho kT$ as a function of $\rho$ for hard spheres (three dimensions) are given in figure~1 when the series is terminated after
$\epsilon_1$, $\epsilon_3$,
and $\epsilon_4$. The results for the case where only $\epsilon_1$ is included because this is Rennert expression.  In this case, the
pressure is monotonous. Results for the cases when the series terminates after $\epsilon_2$ are not included because the pressure is
monotonous.  The displayed curves are compared with the results of computer simulations taken from \cite{2.}.  The simulation curves consist
of two branches, one for the fluid branch and one for the solid branch, which terminates at close packing.  The agreement is quite
reasonable.  However, the most
intriguing feature is not the numerical accuracy of this procedure that, for the fluid branch, is not quite as good as that of the
{\it ad hoc} Carnahan-Starling \cite{3.}
expression but the fact that with terms through $\epsilon_3$ or $\epsilon_4$ included in equation~(12), there is a maximum in the
pressure at approximately the location of the transition
from the fluid to the solid phases.  Better results are given with $\epsilon_4$ included but neglecting the contribution of $\epsilon_4$
does a better job of locating the transition.   It is reasonable to regard this maximum as indicative of the freezing of the hard  sphere
fluid, since a system with a negative compressiblity would expand with an increase of pressure, which is unphysical.
The Carnahan-Starling expression does not give any indication of the presence of this transistion and,
in fact, is unphysical at very high densities since it continues to give results for densities past close packing.  To be sure, the procedure
reported here is {\it ad hoc}.
There is no guarantee that including more $\epsilon_i$ will continue to give better results or even a maximum pressure.  In fact, this is
the situation in two dimensions where neglecting $\epsilon_4$ yields a maximum while including $\epsilon_4$ does not.  The point of this
article is to point out that this procedure does give an indication of a phase transition
in a hard sphere fluid.  To the author's knowledge, beyond the more complex Born-Green-Yvon \cite{4.,4._2,5.} and Kirkwood \cite{6.} approximations, this is the only theory to do so.  This paper might probably point the way to a simple non-empirical description of the hard sphere transition and
be useful in this regard.

\section*{Acknowledgements}

It is a pleasure to enthusiastically acknowledge the lifetime achievements by Myroslav Holovko and my friendship with him.

\ukrainianpart

\title{Немонотонний тиск як функція густини у плині без притягальних сил}

\author{Д. Гендерсон}
\address{Відділ хімії та біохімії, Університет Брайхем Янг, Прово, США}

\makeukrtitle
\begin{abstract}
Простий результат для тиску плину твердих сфер, який був отриманий
багато років тому Реннертом, розширено в простий спосіб шляхом
додавання членів, які мають такий же вигляд як формула Реннерта.
Результуючий вираз є посередньо точним, але його точність  не
обов'язково  покращиться, якщо включити додаткові члени. Цікавим
наслідком отриманого виразу є те, що тиск може мати максимум, коли
густина зростає, що узгоджується із твердненням твердих сфер. Це
відбувається виключно як наслідок короткодійних взаємодій. Лише
теорії Борна-Гріна-Івона і Кірквуда показують таку поведінку для
твердих сфер і вони потребують числового розв'язку інтегрального
рівняння. Процедура, окреслена тут є {\it ad hoc}, але можливо є
корисною такою ж мірою, як і популярне рівняння Карнагана-Старлінга
для тиску твердих сфер, яке є також {\it ad hoc}, але корисним.

\keywords статистична сума, рівняння стану, тиск, плин твердих сфер,
перехід тверднення

\end{abstract}

\end{document}